# Isostructural Metal-Insulator Transition Driven by Dimensional-Crossover in SrIrO₃ Heterostructures

*Shuai Kong,[1, 2, 3, ‡] Lei Li,[1, 2, 3, ‡] Zengxing Lu,[1, 2, 3, ‡] Jiatai Feng,[1, 2, 3] Xuan Zheng,[1, 3, 4] Pengbo Song,[5] You-guo Shi,[5] Yumei Wang,[5] Binghui Ge,[6] Katharina Rolfs,[7] Ekaterina Pomjakushina,[7] Thorsten Schmitt,[8] Nicholas C. Plumb,[8] Ming Shi,[8] Zhicheng Zhong,[1, 2, 3, *] Milan Radovic,[8, 9, *] Zhiming Wang,[1, 2, 3, *] and Run-Wei Li[1, 2, 3]*

[1]Key Laboratory of Magnetic Materials and Devices, Ningbo Institute of Materials Technology and Engineering, Chinese Academy of Sciences, Ningbo 315201, China

[2]Center of Materials Science and Optoelectronics Engineering, University of Chinese Academy of Sciences, Beijing 100049, China

[3]Zhejiang Province Key Laboratory of Magnetic Materials and Application Technology, Ningbo Institute of Materials Technology and Engineering, Chinese Academy of Sciences, Ningbo 315201, China

[4]Department of Chemical and Environmental Engineering, The University of Nottingham, Ningbo 315042, China

[5]Institute of Physics, Chinese Academy of Sciences, Beijing 10084, China

[6]Institutes of Physical Science and Information Technology, Anhui University, Hefei 230601, China

[7]Laboratory for Multiscale Materials Experiments, Paul Scherrer Institut, CH-5232 Villigen PSI, Switzerland

[8]Swiss Light Source, Paul Scherrer Institut, CH-5232 Villigen PSI, Switzerland

[9]SwissFEL, Paul Scherrer Institut, CH-5232 Villigen PSI, Switzerland

[‡] These authors contributed equally to this work

[*]e-mail: zhiming.wang@nimte.ac.cn; milan.radovic@psi.ch; zhong@nimte.ac.cn

**Abstract:** Dimensionality reduction induced metal-insulator transitions in oxide heterostructures are usually coupled with structural and magnetic phase transitions, which complicate the interpretation of the underlying physics. Therefore, achieving isostructural MIT is of great importance for fundamental physics and even more for applications. Here, we report an isostructural metal-insulator transition driven by dimensional-crossover in spin-orbital coupled $SrIrO_3$ films. By using *in-situ* pulsed laser deposition and angle-resolved photoemission spectroscopy, we synthesized and investigated the electronic structure of $SrIrO_3$ ultrathin films with atomic-layer precision. Through inserting orthorhombic $CaTiO_3$ buffer layers, we demonstrate that the crystal structure of $SrIrO_3$ films remains bulk-like with similar oxygen octahedra rotation and tilting when approaching the ultrathin limit. We observe that a dimensional-crossover metal-insulator transition occurs in isostructural $SrIrO_3$ films. Intriguingly, we find the bandwidth of $J_{eff}=3/2$ states reduces with lowering the dimensionality and drives the metal-insulator transition. Our results establish a bandwidth controlled metal-insulator transition in the isostructural $SrIrO_3$ thin films.

# I. INTRODUCTION

Oxide heterostructures offer unprecedented opportunities to tune the interactions with similar energy scales, e.g., crystal field splitting, electron correlation and spin-orbit coupling for modulating the ground state of transition metal oxides [1-6]. Among all tuning knobs, dimensionality engineering has emerged as an effective approach to control magnetic and electronic properties in oxides heterostructures in a layer-by-layer manner thanks to the tremendous development of oxide thin film growth [7]. As a prominent example, the metal-insulator transition (MIT) is widely observed in oxide heterostructures through reducing dimensionality [8-15]. However, the dimensional-crossover driven MIT is commonly accompanied by an associated structural phase transition since transition metal oxides exhibit strong coupling between charge, spin, and lattice degrees of freedom [16]. This coexistence obscures the underlying physics, making it difficult to disentangle the different intrinsic interactions that control the MIT. Therefore, achieving an isostructural MIT through dimensional-crossover and consequently controlling physical properties of oxide heterostructures is of great fundamental and technological interests [17,18].

Recently, the $5d$ transition metal iridates emerge as an ideal platform for exploring novel quantum states of matter due to the interplay between strong spin-orbit coupling and electron correlations [5,6,19-21]. Among all iridates, the three-dimensional perovskite $SrIrO_3$ has attracted tremendous attention as a key building block for engineering electronic, magnetic and topological phases in oxide heterostructure [22-28]. The bulk $SrIrO_3$ exhibits paramagnetic and semimetal behavior with proposed Dirac nodal line and large spin Hall effect, while being at the border of the MIT [28-30]. Through reducing dimensionality, $SrIrO_3$ film undergoes a MIT, accompanied by dramatic changes of the magnetic and structural phases [16,31-33]. Intriguingly, the crystal structure transforms from orthorhombic to tetragonal upon reducing $SrIrO_3$ thickness, where the oxygen octahedra tilting along [001] direction is suppressed. Such entangled multiple phase transitions have been also reported in Ruddlesden-Popper series [$(SrIrO_3)_n$, SrO] [12] and mimic two-dimensional layered [$SrIrO_3$,$SrTiO_3$] superlattices [13], complicating the understanding the underlying mechanism of the dimensional-crossover driven MIT. Moreover, the structural degree of freedom in iridates plays an important role in the electronic and magnetic properties [34]. To isolate the structural and electronic contribution of the dimensional-crossover driven MIT, it is of great importance to systematically investigate the electronic structure evolution in $SrIrO_3$ ultrathin films, while the structural degree of freedom is kept intact.

In this letter, we demonstrate an isostructural MIT driven by dimensional-crossover in spin-orbital coupled $SrIrO_3$ films. By controlling oxygen octahedral connectivity via inserted orthorhombic $CaTiO_3$ buffer layers, we maintain $SrIrO_3$ ultrathin films in the orthorhombic crystal structure with bulk-like oxygen octahedra rotation and tilting during reducing film thickness. Employing in-situ angle-resolved photoemission spectroscopy (ARPES) we track the evolution of the electronic structure as a function of $SrIrO_3$ film thickness. Intriguingly, we identify that the bandwidth of $J_{eff}=3/2$ states shows negligible change in thicker films and starts to decrease when the film thickness is below 4 unit cells (u.c.), leading to an isostructural dimensional-crossover driven MIT in $SrIrO_3$ ultrathin films.

## II. RESULTS AND DISCUSSION

Epitaxial $SrIrO_3/CaTiO_3$ heterostructures (see Figure 1(a)), consisting of varying $SrIrO_3$ thickness between 10 to 1 u.c. and fixing $CaTiO_3$ thickness at 5 u.c. as buffer layers, were synthesized by pulsed laser deposition (PLD) on (001) $TiO_2$-terminated Nb-doped (0.5 w%) $SrTiO_3$ substrates. The film growth was monitored with atomic u.c. precision by using high-pressure reflection high energy electron diffraction (RHEED). The surface structure of thin films was verified by low energy electron diffraction (LEED). After immediately transferring the sample into the ARPES chamber under ultrahigh vacuum condition, in-situ ARPES measurements were performed at the SIS beamline in Swiss Light Source. First-principles density functional theory (DFT) calculations were performed with the generalized gradient approximation using the plane wave VASP package. More details regarding to the growth, sample characterization and DFT calculations can be seen in the section of METHODS.

Figure 1 summarizes the control of oxygen octahedral out-of-plane tilting in $SrIrO_3$ heterostructures through interface engineering. The orthorhombic $SrIrO_3$ possesses an $a^+b^-b^-$ rotation in Glazer notation [35]. In contrast, the cubic $SrTiO_3$ substrate possesses an $a^0a^0a^0$ rotation without in-plane rotation and out-of-plane tilting at room temperature. Strong mismatch of oxygen octahedra occurs and results in suppression of the out-of-plane tilting in $SrIrO_3$ ultrathin films when grown on the $SrTiO_3$ substrate [16,32,33]. As shown in Ref. 16, the structural transition from orthorhombic to tetragonal during reducing film thickness can be inferred from LEED measurements. This finding is in good agreement with our density functional theory (DFT) calculations shown in Figure 1(b)-1(d) (details of calculations are described in the section of METHODS). Intriguingly, after inserting orthorhombic CaTiO$_3$ buffer layers, it is found that both oxygen octahedral out-of-plane tilting and in-plane rotation in SrIrO$_3$ are preserved even in the ultrathin limit. The presence of the out-of-plane tilting and in-plane rotation enlarge surface periodicity as illustrated in Figure 1(e), giving rise to a $2 \times 2$ LEED pattern in Figure 1(f)-1(h) [16]. Moreover, the $2 \times 2$ LEED patterns are present for all the SrIrO$_3$/CaTiO$_3$ heterostructures with varying SrIrO$_3$ thickness from 10 to 1 u.c. (seen in Supplemental Material Figure S1). Together, our theoretical and experimental results conclusively demonstrate that orthorhombic-like SrIrO$_3$ structure has been stabilized without structural transition during reducing thickness after introducing the CaTiO$_3$ buffer layers.

Previous studies have found a concurrent metal-insulator and structural transitions as SrIrO$_3$ film thickness decreases [16,31]. We track how the electronic structure evolves while maintaining the crystal structure intact during reducing SrIrO$_3$ film thickness down to the ultrathin limit. Figure 2(a) gives the three-dimensional Brillouin zone of orthorhombic SrIrO$_3$. Figure 2(b) shows the Fermi surface intensity map of 10 u.c. SrIrO$_3$ grown on CaTiO$_3$ buffer layer, which displays hole and electron pockets around $\Gamma$, S and X points, respectively. The Fermi surface is consistent with the semi-metal character found in thick SrIrO$_3$ thin films [30,36]. Figure 2(c) show the energy momentum maps along $\Gamma$-X direction of SrIrO$_3$/CaTiO$_3$ heterostructures with varying SrIrO$_3$ thickness from 10, 5, 4, 3, 2 down to 1 u.c. and CaTiO$_3$ buffer layer fixed at 5 u.c.. The band dispersion shows negligible change when the thickness > 3 u.c., however, the spectral weight near E$_F$ is gradually suppressed and transferred into higher binding energy during further reducing film thickness, and eventually an energy gap is clearly opened in 2 and 1 u.c. SrIrO$_3$ ultrathin films.

To further characterize the isostructural MIT, we focus now on the evolution of energy distribution curves shown in Figure 2(d). The coherent quasi-particle peaks around E$_F$ diminish and transfer into incoherent peaks as the film thickness decreases. To describe this quantitatively, we use an effective quasi-particle residue $Z$ by fitting the energy distribution curves (details of the fits are described in the Supplemental Material Figure S2). The quasi-particle residue $Z$ gradually decreases when the film thickness > 3 u.c., below which it suddenly drops nearly to zero. Figure 2(e) shows the comparison of the quasi-particle residue $Z$ and tilting angles as obtained above for SrIrO$_3$ films grown with and without CaTiO$_3$ buffer layers. Similar evolution of $Z$ are observed in both SrIrO$_3$ thin films regardless of CaTiO$_3$ buffer layers, while the evolution of tilting angles differ dramatically. It is clear that the MIT occurs without the structural transition, establishing an isostructural MIT driven by dimensional-crossover in SrIrO$_3$ heterostructures.

In the following, we discuss the mechanism of the dimensional-crossover MIT in isostructural SrIrO$_3$ by taking into account chemical shift and electronic bandwidth ($W$) of the electronic states. We first show the thickness-dependent evolution of Ir 4$f$ core level and O 2$p$ valence band to quantify the chemical shift in Figure 3. Each of the Ir 4$f$ spectra exhibits similar spectral features which can be decomposed into two doublets in Figure 3(a). The doublet feature has been observed previously, corresponding to screened and unscreened components of an Ir$^{4+}$ state [37]. The Ir 4$f$ spectra show negligible shifting in binding energy when thickness > 3 u.c., while turning to higher binding energy upon further reducing film thickness as marked by black lines in Figure 3(a). Accordingly, Figure 3(b) show the spectra of O 2$p$ valence band shifting to higher binding energy (see details of O 2$p$ valence band with hv=82 eV and hv=70 eV in Supplemental Material Figure S3). The relative chemical shift of Ir 4$f$ and O 2$p$ are summarized in Figure 3(c). The cove level and valence band spectra show equivalent shifting of ~0.4 eV to higher binding energy. We note that negligible charge transfer occurs between SrIrO$_3$ and CaTiO$_3$, consistent with recent theoretical calculations [38]. Thus, the observed rigid band shifting can't induce the dimensional-crossover driven MIT.

Next, we track the evolution of the electronic bandwidth W in the isostructural SrIrO$_3$ heterostructures. Figure 4(a)-4(d) show the photoemission intensity and the corresponding second-derivative maps of 10 u.c. and 3 u.c. SrIrO$_3$ with buffer layers, respectively. The strong dispersive band around Γ point away from fermi level about 0.2 eV is identified as the top position of J$_{eff}$=3/2 band as shown in Figure 4(a)-4(c), which is consistent with theoretical calculations and previous ARPES measurements of bulk SrIrO$_3$ [30,36,39,40]. We note that the J$_{eff}$=3/2 band in 3 u.c. thin film shifts significantly to higher binding energy as shown in Figure 4(b)-4(d). Figure 4(e) shows the systematical evolution of J$_{eff}$=3/2 band around Γ point as a function of film thickness (see details in Supplementary Materials Figure S4). The band dispersions are overlaid on each other and remain almost identical when the thickness > 3 u.c., while for thinner films they progressively shift to higher binding energy and finally increase to -1.05 eV. Meanwhile, we note that the bottom position of J$_{eff}$=3/2 band coincides with the onset of the oxygen valence band, consistent with theoretic calculations [41]. The bottom position of J$_{eff}$=3/2 states shows negligible change as the film thickness reduces (see more details in Supplemental Material Figure S3). In Figure 4(f), we summarize the evolution of $J_{eff}=3/2$ band top and bottom position, from which we can determine the $J_{eff}=3/2$ states shifting to higher binding energy and the effective bandwidth narrowing by 0.8 eV during reducing the film thickness.

In Figure 4(g), we present the evolution of $J_{eff}=3/2$ and $J_{eff}=1/2$ states during dimensional-crossover. The observed $J_{eff}=3/2$ band narrowing and shifting to higher binding energy features agree well with the optical spectra in Ruddlesden-Popper series [$(SrIrO_3)_n$, SrO] [12,39,42]. The reduction of bandwidth $W$ is understood as breaking of translation symmetry along the z direction, which in turn provides less hopping channels in the ultrathin films. Moreover, we note that the electron correlation ($U$) monotonically increases as the dimensionality reduces [16,42]. Therefore, the cooperative effect between the reduction of bandwidth $W$ and enhancement of electron correlation $U$ makes the overall effective bandwidth $U/W$ increase dramatically, which in turn triggers the dimensional-crossover driven MIT.

## III. CONCLUSION

In conclusion, we demonstrated that the orthorhombic-like $SrIrO_3$ films have been stabilized without a structural transition upon approaching the ultrathin limit through oxygen octahedra rotation engineering by inserting the orthorhombic $CaTiO_3$ buffer layer. Our momentum-resolved band structure measurements revealed that the dimensional-crossover driven MIT occurs in $SrIrO_3$ ultrathin films regardless of the structural transition, which allows to disentangle the electronic and structural effects in the MIT. We further determined that the reduction of the effective bandwidth plays a crucial role in the dimensional-crossover driven MIT in the isostructural $SrIrO_3$ thin films. Our study provides opportunities for exploring the pure dimensionality effect on the topological phase transitions and spin-orbitronics in isostructural $SrIrO_3$ thin films and heterostructures [28,29].

## ACKNOWLEDGMENTS

This work was supported by the National Basic Research of China (Nos. 2017YFA0303602, 2019YFA0307800, 2017YFA0302901), the National Natural Science Foundation of China (Nos. U1832102, 11874367, 51931011, 51902322, 11774399), the Key Research Program of Frontier Sciences, CAS (No. ZDBS-LY-SLH008), the Thousand Young Talents Program of China, the Natural Science Foundation of Zhejiang province of China (No. LR20A040001), the 3315 Program of Ningbo, the Ningbo Natural Science Foundation (No. 2019A610050, 2019A610055), the Beijing National Laboratory for Condensed Matter Physics. T.S. was supported by the Swiss National Science Foundation (SNSF) through the Sinergia network Mott Physics Beyond the Heisenberg Model (MPBH) (SNSF Research Grants CRSII2_160765/1 and CRSII2_141962). M.R. acknowledges support from SNSF via Research Grant 200021_182695. The authors acknowledge the beamtime on SIS beamline of the Swiss Light Source and Dreamline of the Shanghai Synchrotron Radiation Facility.

The authors declare that they have no conflict of interest.

## APPENDIX: METHODS

**1. Sample growth and characterization:** Epitaxial $SrIrO_3/CaTiO_3$ heterostructures were synthesized on $TiO_2$-terminated Nb-doped (0.5 wt%) $SrTiO_3(001)$ substrates by pulsed laser deposition. The heterostructures consist of 5 unit cells (u.c.) $CaTiO_3$ as a fixed buffer layer, and $SrIrO_3$ layers with thickness varying from 1 u.c. to 10 u.c.. The film growth was monitored with atomic unit cell precision by using high-pressure reflection high energy electron diffraction (RHEED). During growth, the substrate was heated resistively around 750 ℃. The oxygen partial pressure during thin film growth was $1\times10^{-1}$ mbar. After growth, the samples were immediately transferred into the analysis chamber under ultrahigh vacuum condition to perform the low energy electron diffraction (LEED) and angle-resolved photoemission spectroscopy (ARPES) measurements. In-situ ARPES is used to investigate the thickness-dependent evolution of electronic structure of $SrIrO_3$ thin films. ARPES measurements were performed at the SIS beamline in a photon energy range of hv= 70-119 eV. The angular and energy resolutions were better than 0.3 degrees and 20 meV, respectively. All samples were measured around 14 K under a vacuum better than $1\times10^{-10}$ mbar.

**2. Density functional theory calculations:** Our density functional theory calculations were performed with the Vienna ab initio simulation package (VASP). The generalized gradient approximation (GGA) is used to treat the exchange-correlation potential. For the Brillouin zone integration, we use the k meshes of $4\times4\times1$ and the cut off energy for the plane wave of 400 eV for all calculations. We fixed the bottom three layers atoms of the substrate and a conjugate-gradient algorithm is used to relax the other ions. The convergence criterion is set to 0.001 eV between two ionic steps relaxation.

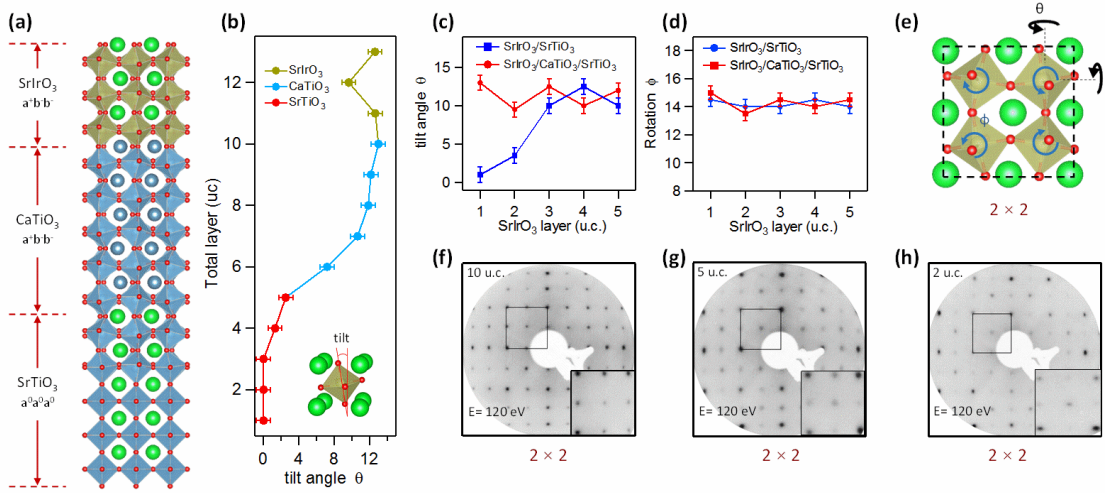

**Figure 1. Control of oxygen octahedral out-of-plane tilting in SrIrO₃/CaTiO₃ heterostructures.** (a) DFT calculated structure of the SrIrO₃/CaTiO₃ heterostructure on Nb:SrTiO₃ substrate. (b) DFT calculated layer-dependent tilting angle of the heterostructure. (c) The octahedral out-of-plane tilting and (d) in-plane rotation angles as a function of film thickness with and without CaTiO₃ buffer layers. The octahedral tilting is roughly independent on the film thickness after interfacing buffer layers. (d) Top view of structural model. The octahedral out-of-plane tilting and in-plane rotation produce a 2 × 2 surface periodicity, which has been observed by LEED pattern in 10 u.c., 5 u.c., 2u.c. SrIrO₃ with CaTiO₃ buffer layers (f)-(h).

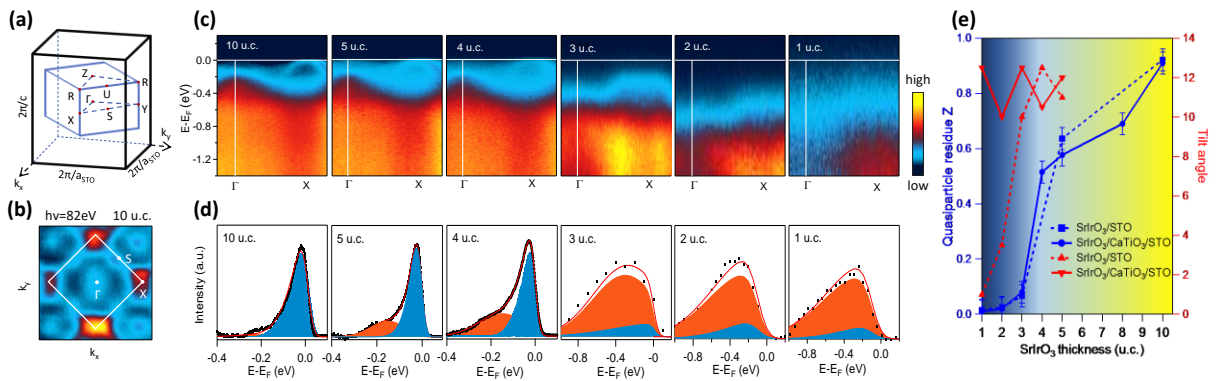

**Figure 2. Dimensional-crossover driven metal-insulator transition in isostructural SrIrO₃.** (a) The three dimensional Brillouin zone (BZ) of orthorhombic SrIrO₃. (b) Fermi surface intensity map measured on 10 u.c. SrIrO₃/CaTiO₃ heterostructure. (c) Energy vs. momentum intensity maps along Γ-X direction with decreasing SrIrO₃ thickness. (d) Energy distribution curves (EDCs) at the Fermi wavevector indicated by the vertical line in the corresponding image plots in (c). An exponential background has been subtracted from the raw EDCs (see Supplemental Material Figure S3). (e) The comparison of quasi-particle residue $Z$ and tilting angles for SrIrO$_3$ films grown with and without CaTiO$_3$ buffer layers. All data were taken at a photon energy of 82 eV with circular polarization and at temperature of 14 K.

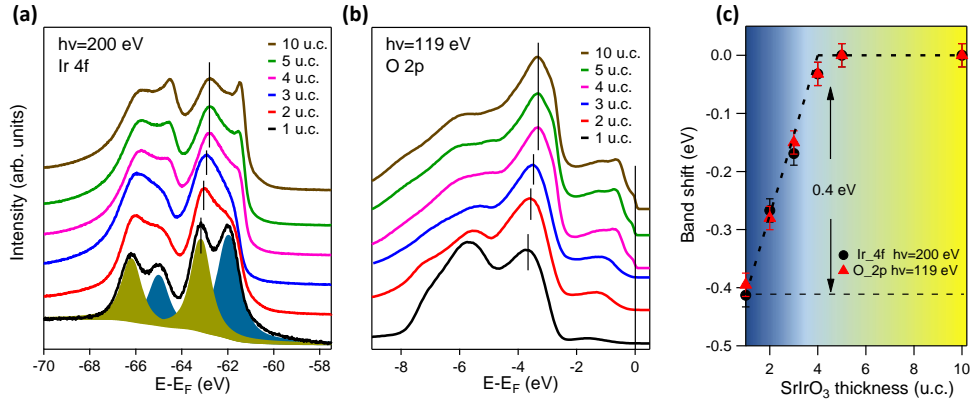

**Figure 3. Thickness-dependent evolution of chemical shift.** The spectra of (a) Ir 4$f$ core-level and (b) O 2$p$ valence band as a function of film thickness. The peak positions are marked with black lines. (c) The relative peak shift of both O 2$p$ and Ir 4$f$ with reference to that of the thick SrIrO$_3$ film.

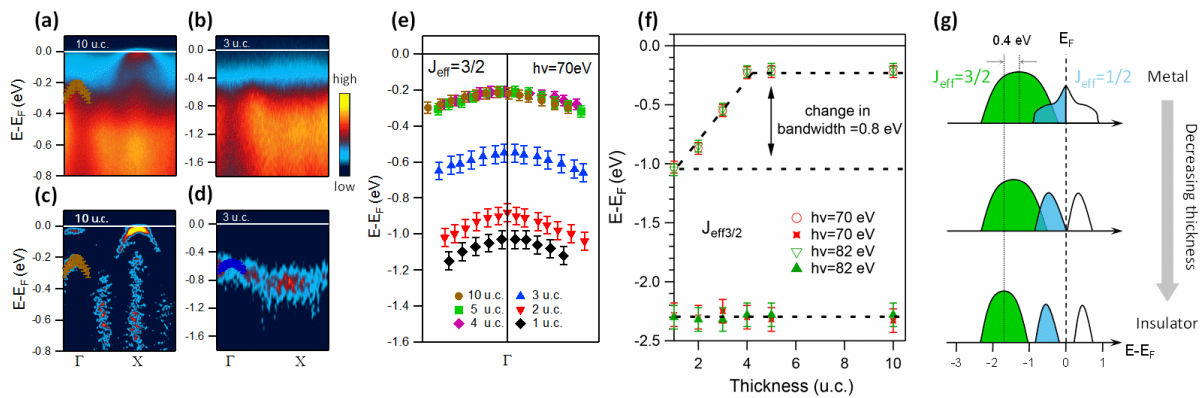

**Figure 4. Mechanism of Dimensional-crossover driven metal-insulator transition in isostructural SrIrO$_3$.** The photoemission intensity plots of (a) 10 u.c. and (b) 3 u.c. SrIrO$_3$ with CaTiO$_3$ buffer layers taken along Γ-X high symmetry direction. (a) and (d) The corresponding

second-derivative plots. (e) Evolution of the $J_{eff} = 3/2$ states around $\Gamma$ point. The data are extracted from (a)-(d) as a function of film thickness. Details are described in the Supplemental Material Figure S4. (f) Evolution of band top and bottom position of the $J_{eff} = 3/2$ states. (g) The thickness-dependent evolution in bandwidth of $J_{eff} = 3/2$ and $J_{eff} = 1/2$ states.

# Supplemental Material

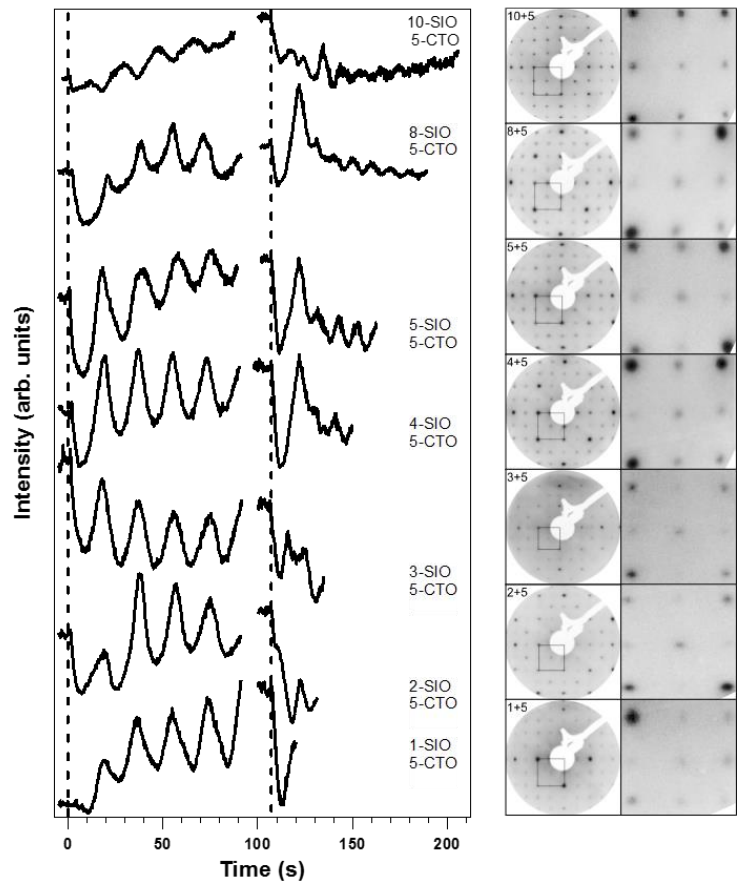

**Figure S1.** The atomic layer-by-layer growth and surface characterization of SrIrO₃ films with CaTiO₃ buffer layers. The left column is RHEED intensity oscillation patterns shown for a series of 1, 2, 3, 4, 5, 8, 10 u.c. SrIrO₃ ultrathin films with 5 u.c. CaTiO₃ buffer layers. The right column is the corresponding LEED patterns recorded with a beam energy of 120 eV. A representative 2×2 surface periodicity structure is indicated by the black squares.

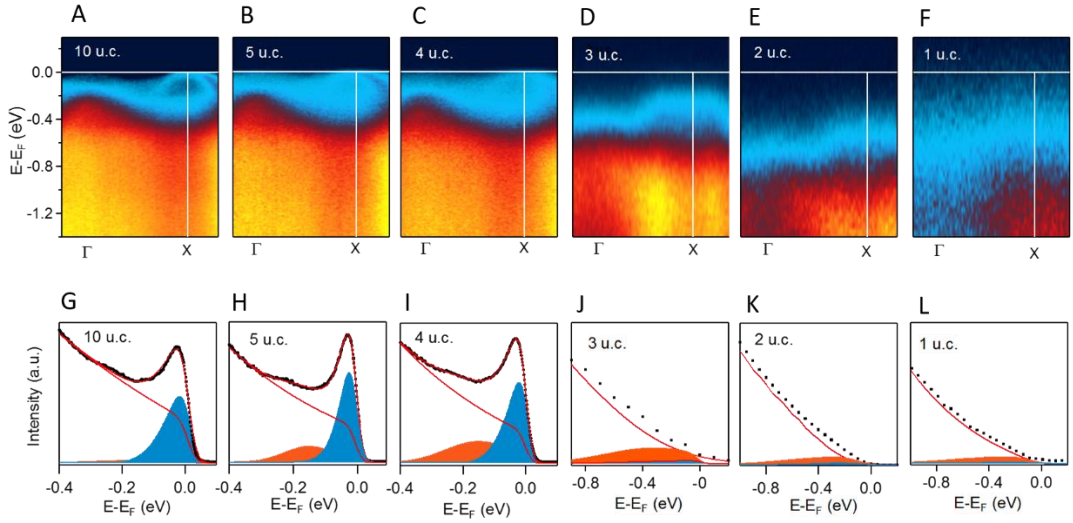

**Figure S2.** Thickness-driven metal-insulator transition in SrIrO₃/CaTiO₃ heterostructures. (a)-(f) Raw energy-momentum intensity maps of different thickness SrIrO₃ along Γ-X. In (g)-(l) we how the results of the raw EDCs in SrIrO₃/CaTiO₃ heterostructures. The black and red lines represent the raw and background data.

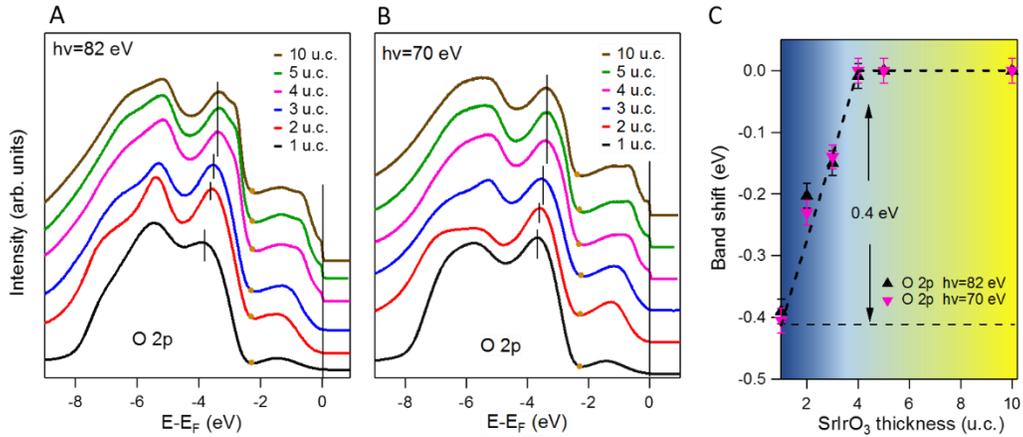

**Figure S3.** Thickness-dependent evolution of O 2p valence bands. The spectra of O 2p valence bands as a function of film thickness with (a) hν=82 eV and (b) 70 eV, respectively. (c) The relative peak shift of O 2p reference to that of thick SrIrO₃ films.

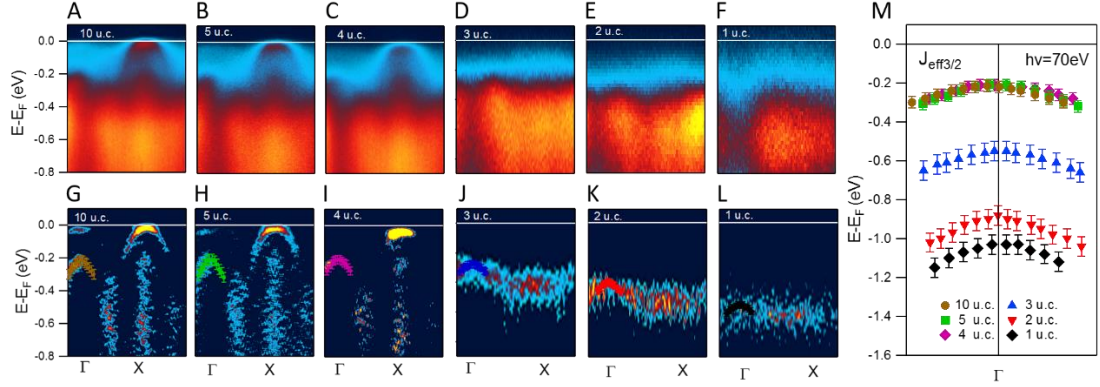

**Figure S4.** Thickness-dependent bandwidth evolution of $J_{eff}=3/2$ states with hv=70 eV. (a)-(f) The photoemission intensity plots of 10 to 1 u.c. $SrIrO_3$ with $CaTiO_3$ buffer layers taken along Γ-X high symmetry direction. (g)-(l) The corresponding second-derivative plots. (m) Evolution of the $J_{eff}=3/2$ states around Γ point.

Figure S2(a)-S2(f) show the energy momentum maps along Γ-X direction of $SrIrO_3$/$CaTiO_3$ heterostructures with varying $SrIrO_3$ thickness from 10, 5, 4, 3, 2 down to 1 u.c. and $CaTiO_3$ buffer layer fixing at 5 monolayers. The coherent quasiparticle peaks around Fermi level ($E_F$) diminish and transfer into incoherent peaks as the film thickness decreases. To quantitatively characterize the loss of quasiparticle coherence, we fit two Lorentzian peaks to the EDCs extracted at the Fermi wavevector. These account for the coherent and incoherent components of the spectral function, as shown in Figure S2(g)-S2(l). From the areas, we define an effective quasiparticle residue as $Z=A_{coh}/(A_{coh}+A_{inc})$ [1,2]. This reflects the reduction of the quasiparticle weight relative to that bulk, the quantity of interest as we reduce the film thickness down to nanoscale dimensions. The Ir 5d derived states near the $E_F$ almost unchange down to 4 MLs, and consist of the coherent part (quasiparticle peak) located at $E_F$. In 3 MLs, the spectral weight starts to transfer from the coherent part to the incoherent part. The gap is fully opened when the film thickness down to 2-1 MLs. The loss in quasiparticle coherent at 3 u.c. indicates the corresponding thickness-driven metal-insulator transition. We have found that the MIT of $SrIrO_3$ during the lowering of dimension still exists after inserting orthorhombic $CaTiO_3$ buffer layers.

We show the thickness-dependent evolution of O 2p valence band with hv=82 eV and hv=70 eV shown in Figure S3(a) and S3(b). Accordingly, the relative chemical shift of O 2p are summarized in Figure S3(c). The O 2p spectra show equivalent shift of ~0.4 eV to higher binding energy, indicating rigid band shifting and negligible charge transfer between SrIrO₃ and CaTiO₃, in consistent with recent theoretical calculations [3].

Figures S4(a)-S4(f) and S4(g)-S4(l) show the photoemission intensity and the corresponding second-derivative maps of 10, 5, 4, 3, 2, 1 u.c. SrIrO₃ with buffer layers, respectively. The strong dispersive band around Γ point away from fermi level about 0.2 eV is identified as the top position of $J_{eff}$=3/2 band, which is consistent with theoretical calculation and previous ARPES measurement of bulk SrIrO₃ [4,5]. We note that the $J_{eff}$=3/2 band in 3, 2, 1 u.c. thin film shifts significantly to higher binding energy. Figure S4 summarize the systematical evolution of $J_{eff}$=3/2 band around Γ point as a function of film thickness. The band dispersion are overlayed each other and remain almost same when the thickness > 3 u.c., below which it gradually shifts to higher binding energy and finally increase to -1.05 eV.